\documentclass[12pt]{article} 
\oddsidemargin = \evensidemargin 
\newfont{\sase}{cmss12 scaled 1300} 
\newfont{\sasi}{cmss12 scaled 1500} 
\usepackage{colordvi} 
\usepackage{epsfig} 
\usepackage{amssymb}

\begin{document} 

\title{\bf Complexity and Non-Commutativity \\ of  
Learning Operations on Graphs} 

\bigskip 

\author{{\bf Harald Atmanspacher and Thomas Filk$^*$}\\  \\ 
Institute for Frontier Areas of Psychology and Mental Health, \\ 
Wilhelmstr.~3a, 79098 Freiburg, Germany   
\and   
Parmenides Foundation, \\ 
Via Mellini 26--28, 57031 Capoliveri, Italy 
\and 
$^*$ also at Institut f\"ur Physik, Universit\"at Freiburg, \\ 
Hermann--Herder--Str.~3, 79104 Freiburg, Germany} 

\date{} 
\maketitle 

\bigskip 

\vskip 4cm

\begin{abstract} 

We present results from numerical studies of supervised learning operations in recurrent networks 
considered as graphs, leading from a given set of input conditions to predetermined outputs. Graphs 
that have optimized their output for particular inputs with respect to predetermined 
outputs are asymptotically stable and can be characterized by attractors which form a 
representation space for an associative multiplicative structure of input operations. As the 
mapping from a series of inputs onto a series of such attractors generally depends on 
the sequence of inputs, this structure is generally non-commutative. Moreover, the 
size of the set of attractors, indicating the complexity of learning, is found to behave 
non-monotonically as learning proceeds. A tentative relation between this complexity and 
the notion of pragmatic information is indicated.
  
\end{abstract} 


\newpage 

\section{Introduction} 

Graph theory has recently reveived increasing attraction for applications to 
complex systems in various disciplines (Gernert 1997, Paton 2002a,b, 
Bornholdt and Schuster 2003). The characterization of systems 
(with interrelated constituents) by graphs (with linked vertices) is comparably 
general as their characterization in terms of categories 
(with elements related by morphisms). Despite its generality, 
graph theory has turned out to be a powerful tool for gaining very specific 
insight into structural and dynamical properties of complex systems (see 
Jost and Joy 2002, Atmanspacher et al.~2005 for examples).        

An area of particularly intense interest, in which complex systems abound, is 
biological information processing. This ranges from evolutionary biology 
over genetics to the study of neural systems. Theoretical and computational 
neuroscience have become rapidly growing fields (Hertz et al. 1991, Haykin 1999, 
Dayan and Abbott 2001) in which graph theoretical methods have gained considerable 
significance (cf.~Sejnowski 2001). 

Two basic classes of biological networks are feedforward and recurrent
networks. In networks with purely feedforward (directed) connectivities, neuronal input is 
mapped onto neuronal output through a feedforward synaptic weight matrix. 
In recurrent networks, there are additional (directed or bi-directed) connectivities 
between outputs and other network elements, giving rise to a recurrent synaptic weight matrix.  
Much recurrent modeling incorporates the theory of nonlinear and complex dynamical systems
(cf.~Smolensky 1988, see also beim Graben 2004 for discussion).

Hopfield networks are an example of a fully recurrent network 
in which all connectivities are bidirectional and the output is a deterministic 
function of the input. Their stochastic generalizations are known as Boltzmann machines. 
Another important distinction with respect to the implementation of neural networks
refers to the way in which the neuronal states are characterized:
the two main options are firing rates and action potentials (for more details see
Haykin 1999).
     
A key topic of information processing in complex biological networks is learning, 
for which three basically different 
scenarios are distinguished in the literature (see Dayan and Abbott 2001, Chap.~III):   
unsupervised, supervised and reinforcement learning. In unsupervised (also self-supervised) 
learning a network responds to inputs solely on the basis of its intrinsic structure 
and dynamics. A network learns by evolving into a state that is constrained by its own properties 
and the given inputs, an important modelling strategy for implicit learning processes. 

In contrast, supervised learning presupposes the definiton of desired input-output relations, 
so the learned state of the network is additionally constrained by its outputs. Usually, the learning 
process in this case develops by minimizing the difference between the actual output and the 
desired output. The corresponding optimization procedure is not intrinsic to the evolution of the 
system itself, but has to be externally arranged, hence the learning is called supervised. 
If the supervision is in some sense ``naturalized'' by coupling a network to an environment, 
which provides evaluative feedback, one speaks of reinforcement learning. 

In this contribution we are interested in supervised learning (see Duda et al.~2000 
for a review) on small, fully recurrent networks implemented on graphs (cf.~Jordan 1998). 
We start with a general formal characterization in terms of dynamical systems (Sec.~2.1), describe 
how they are implemented on graphs (Sec.~2.2), and show how it reaches asymptotically 
stable states (attractors) when the learning process is terminated, i.e.~is optimized for 
given inputs and (random) initial conditions with respect to predetermined outputs (Sec.~2.3). 

We shall characterize the learning operations by a multiplicative structure characterizing 
successively presented inputs in Sec.~3.1. In this context we confirm and specify earlier 
conjectures (e.g., Gernert 1997) about the non-commutativity of learning operations for a 
concrete model. In Sec.~3.2, we study how the size of the set of attractors representing 
the derived structure changes during the process for perfectly and imperfectly optimized 
networks. The number of attractors is proposed to indicate the complexity of learning, 
and in Sec.~4 this is tentatively related to pragmatic information as a particular measure
of meaning.

\section{Supervised Learning in Recurrent Networks}

\subsection{General Notation} 

Let $M$ be a set, and let $M=X\cup B$, with 
$X\cap B=\emptyset$, be a partition of $M$ into two 
disjoint subsets. If $M$ is some closed subset of 
${\bf R}^n$, $B$ may be the boundary 
of $M$. (Later we will specify $M$ as the 
vertices of a graph, $B$ as a set of ``external'' or ``boundary'' 
vertices, and $X$ as a set of ``internal'' vertices.) 

We consider the dynamics of fields $u(x,y,t)\in U$, 
where $x\in X$, $y\in B$, $t$ represents time as parametrized discretely 
or continuously, and $U$ is the space of admissible state values for the 
fields. The dynamics of $u$ can be described by an equation 
\begin{equation} 
\label{eq1} 
             F[u(x,y,t)]=0  \, . 
\end{equation} 
For a continuous time variable and $M\subset {\bf R}^n$, a typical example 
is the diffusion equation 
\begin{equation} 
    F[u(x,t)]= \frac{\partial u(x,t)}{\partial t} 
         - \lambda \Delta u(x,t)   
\end{equation} 
where $\Delta$ is the Laplace operator and $\lambda$ the diffusion constant. 
The only constraint on Eq.~\ref{eq1} is 
that a state $u(x,y,0)$ at time $t=0$ determines uniquely the solution for any 
time $t>0$. 

We now define a set of external conditions 
$\{b_i:B\rightarrow U\}$ specifying field values $b_i$ on $B$ 
which will be kept fixed during the time evolution 
of the fields on $M$. This is to say that the 
dynamics of fields is effectively restricted to $X$: 
\begin{equation} 
\label{eq2} 
     F[u(x,b_i,t)]=0\, . 
\end{equation}   

Since the state of the system at time $t=0$ uniquely determines 
the states for all $t>0$, we can define a mapping $\Phi_t$, the 
so-called time evolution operator, acting on the set 
of field states. For an initial state $u$ at $t=0$, $\Phi_t[u]$ 
yields the state of the system at $t>0$. Taking into account that 
different external conditions initiate different evolutions, we have 
to specify the time evolution operator as 
a mapping $\Phi_{t,b_i}:F_X\rightarrow F_X$, 
where $F_X$ is the set of states $u(x): X\rightarrow U$, 
by the following construction: 
Let $u(x, t=0)$ be the initial condition for 
Eq.~(\ref{eq2}), then 
\[   \Phi_{t,b_i}[u](x) = u(x,b_i,t) \] 
is the state of the corresponding solution 
at time $t$ under the external condition $b_i$. 

In principle, the state space of $\Phi_{t,b_i}$ can be the entire 
set of states $F_X$. However, for reasons which will become clear below, 
we are interested in dissipative systems evolving into attractors $a_i$ 
in the limit of large $t$. If one of the states belonging to an attractor is 
chosen as an initial condition, the image 
of $\Phi_{t,b_i}$ will again be one of the attractor states. 
This allows us to reduce the 
number of possible states on which the 
mappings $\Phi_{t,b_i}$ close. 

Denoting the flow operator $B_i\equiv \Phi_{t,b_i}$ as the input under the external condition 
$b_i$, we now consider the set of states $A\in F_X$ belonging to attractors after time $t$. 
Then all mappings $B_i$, applied to an attractor $a$, lead to images in $A$: 
\begin{equation} 
\label{eq3} 
  B_i[a] \in A   \hspace{1cm} 
     \mbox{for } a\in A  \, . 
\end{equation} 
In general, the set of all attractor states $A$ does not contain 
a proper subset which is 
mapped onto itself by the set of mappings $\{B_i\}$; otherwise $A$ can be reduced to such a
subset. Each single mapping $B_i$ may not be surjective, 
but the union of the images of all $\{B_i\}$ equals $A$. 

Due to condition (\ref{eq3}), we can define a composition 
of mappings $B_i$. In this way, the external 
conditions $\{b_i\}$ give rise to an associative 
multiplicative structure $\{B_i\}$. This structure is represented on 
the set of attractors  $\{a_i\}$.   

\subsection{Implementation on Graphs} 

We now implement the general notions developed so far on graphs 
(see Wilson 1985 for an introduction to graph theory) and 
specify the set $M$ as the set of vertices $V$ of a 
graph. For simplicity we consider directed graphs with 
single connections for each direction between any two vertices and 
without self-loops. Such a graph gives rise to 
non-reflexive relations on $V$ and can be represented 
by an adjacency matrix $Ad$. 
For two vertices $x_1$ and $x_2$ we have: 
\begin{equation} 
 {Ad}(x_1,x_2) = \left\{ \begin{array}{ll} 
  1 & \mbox{if there exists a directed line from } x_2 
      ~{\rm to}~x_1 \\ 
 0 & \mbox{otherwise} \end{array} \right. 
\end{equation} 
If $Ad$ is symmetric, the graph is undirected. 

The set of vertices $V$ is decomposed into a set 
of external vertices $V_{\rm ext}$ and a 
set of internal vertices $V_{\rm int}$. 
If $N$ is the total number of vertices, 
$N_{\rm ext}$ the number of external vertices and 
$N_{\rm int}$ the number of internal vertices, we 
have $N=N_{\rm ext}+N_{\rm int}$. 

Next we consider fields $u(z,t)$ on a graph with vertices 
$z\in V$ evolving in discrete time steps $t\in {\bf N}$ 
according to: 
\begin{equation} 
\label{eqdyn} 
 u(z,t+1) = f\left( \sum_{y\rightarrow z} u(y,t)\right) 
  = f \left( \sum_y Ad (z,y)u(y,t) \right) \, . 
\end{equation} 
The value of the field $u$ at vertex $z$ and time 
$t+1$ depends only on the sum of the field values at 
neighboring vertices $y$ at time $t$. 

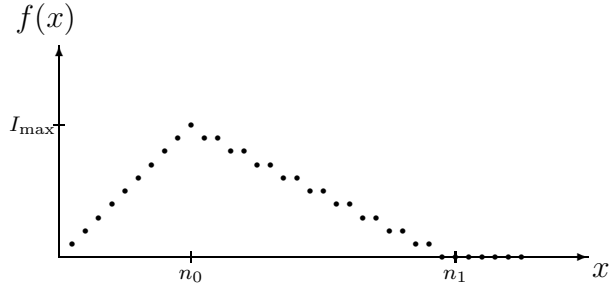
\begin{figure}[htb]
\begin{center}
\begin{picture}(240,100)(0,0)
\put(10,10){\vector(1,0){200}}
\put(10,10){\vector(0,1){80}}
\put(60,8){\line(0,1){4}}
\put(160,8){\line(0,1){4}}
\put(8,60){\line(1,0){4}}
\multiput(15,15)(5,5){10}{\circle*{2}}
\multiput(70,55)(10,-5){10}{\circle*{2}}
\multiput(65,55)(10,-5){10}{\circle*{2}}
\multiput(165,10)(5,0){5}{\circle*{2}}
\put(215,5){\makebox(0,0){$x$}}
\put(60,3){\makebox(0,0){${\scriptstyle n_0}$}}
\put(160,3){\makebox(0,0){${\scriptstyle n_1}$}}
\put(0,60){\makebox(0,0){${\scriptstyle I_{\rm max}}$}}
\put(5,100){\makebox(0,0){$f(x)$}}
\end{picture}
\end{center}
\caption{\small The function $f(x)$ according to Eq.~(\ref{eqf}). The values
$I_{\rm max}=10$, $n_0=10$, and $n_1=30$ are used in the simulations.}
\label{fx}
\end{figure}

The fields $u(z,t)$ assume integer values 
$\{0,1,\ldots,I_{\rm max}\}$, and the function $f$ is defined as: 
\begin{equation} 
\label{eqf} 
   f(x) = \left\{ \begin{array}{ll} 
   {\rm int} \big( I_{\rm max}\cdot(x/n_0)\big) & {\rm for~} x<n_0 \\ 
   {\rm int} \big( I_{\rm max}\cdot(n_1-x)/(n_1-n_0) \big) & {\rm for~} 
    n_0\leq x < n_1 \\ 
   0 & {\rm for~} x\geq n_1 \end{array} 
  \right.    
\end{equation}   
where ${\rm int}(x)$ denotes the nearest integer-rounded $x$. 
The function $f(x)$ is shown in Fig.~\ref{fx}. 

The restriction of $u(z,t)$ to integer values implies that 
there is only a finite number of states. Starting from an arbitrary 
initial state in $F_X$, the system runs   
into an attractor after a few time steps. In many cases, this attractor 
is a fixed point, i.e.~one single state that is asymptotically 
stable. Sometimes the attractor is a limit cycle, i.e.~a periodic 
succession of several (usually few) states. 
Strange attractors do not occur since the number of states is finite. 


The external conditions $\{b_i\}$ are defined as fixed states 
on the external vertices, i.e., the state values on the external vertices 
remain unaffected by the dynamics. Of course, the external conditions 
are supposed to influence the dynamics of the internal vertices. 

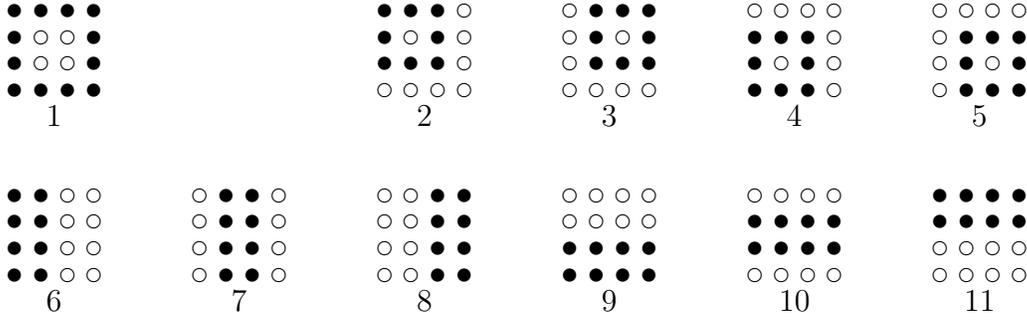
\begin{figure}
\begin{picture}(390,120)(-20,0)
\put(0,80){\circle*{5}}
\put(10,80){\circle*{5}}
\put(20,80){\circle*{5}}
\put(30,80){\circle*{5}}
\put(0,90){\circle*{5}}
\put(10,90){\circle{5}}
\put(20,90){\circle{5}}
\put(30,90){\circle*{5}}
\put(0,100){\circle*{5}}
\put(10,100){\circle{5}}
\put(20,100){\circle{5}}
\put(30,100){\circle*{5}}
\put(0,110){\circle*{5}}
\put(10,110){\circle*{5}}
\put(20,110){\circle*{5}}
\put(30,110){\circle*{5}}
%
\put(140,80){\circle{5}}
\put(150,80){\circle{5}}
\put(160,80){\circle{5}}
\put(170,80){\circle{5}}
\put(140,90){\circle*{5}}
\put(150,90){\circle*{5}}
\put(160,90){\circle*{5}}
\put(170,90){\circle{5}}
\put(140,100){\circle*{5}}
\put(150,100){\circle{5}}
\put(160,100){\circle*{5}}
\put(170,100){\circle{5}}
\put(140,110){\circle*{5}}
\put(150,110){\circle*{5}}
\put(160,110){\circle*{5}}
\put(170,110){\circle{5}}
%
\put(210,80){\circle{5}}
\put(220,80){\circle{5}}
\put(230,80){\circle{5}}
\put(240,80){\circle{5}}
\put(210,90){\circle{5}}
\put(220,90){\circle*{5}}
\put(230,90){\circle*{5}}
\put(240,90){\circle*{5}}
\put(210,100){\circle{5}}
\put(220,100){\circle*{5}}
\put(230,100){\circle{5}}
\put(240,100){\circle*{5}}
\put(210,110){\circle{5}}
\put(220,110){\circle*{5}}
\put(230,110){\circle*{5}}
\put(240,110){\circle*{5}}
%
\put(280,80){\circle*{5}}
\put(290,80){\circle*{5}}
\put(300,80){\circle*{5}}
\put(310,80){\circle{5}}
\put(280,90){\circle*{5}}
\put(290,90){\circle{5}}
\put(300,90){\circle*{5}}
\put(310,90){\circle{5}}
\put(280,100){\circle*{5}}
\put(290,100){\circle*{5}}
\put(300,100){\circle*{5}}
\put(310,100){\circle{5}}
\put(280,110){\circle{5}}
\put(290,110){\circle{5}}
\put(300,110){\circle{5}}
\put(310,110){\circle{5}}
%
\put(350,80){\circle{5}}
\put(360,80){\circle*{5}}
\put(370,80){\circle*{5}}
\put(380,80){\circle*{5}}
\put(350,90){\circle{5}}
\put(360,90){\circle*{5}}
\put(370,90){\circle{5}}
\put(380,90){\circle*{5}}
\put(350,100){\circle{5}}
\put(360,100){\circle*{5}}
\put(370,100){\circle*{5}}
\put(380,100){\circle*{5}}
\put(350,110){\circle{5}}
\put(360,110){\circle{5}}
\put(370,110){\circle{5}}
\put(380,110){\circle{5}}
%
\put(0,10){\circle*{5}}
\put(10,10){\circle*{5}}
\put(20,10){\circle{5}}
\put(30,10){\circle{5}}
\put(0,20){\circle*{5}}
\put(10,20){\circle*{5}}
\put(20,20){\circle{5}}
\put(30,20){\circle{5}}
\put(0,30){\circle*{5}}
\put(10,30){\circle*{5}}
\put(20,30){\circle{5}}
\put(30,30){\circle{5}}
\put(0,40){\circle*{5}}
\put(10,40){\circle*{5}}
\put(20,40){\circle{5}}
\put(30,40){\circle{5}}
%
\put(70,10){\circle{5}}
\put(80,10){\circle*{5}}
\put(90,10){\circle*{5}}
\put(100,10){\circle{5}}
\put(70,20){\circle{5}}
\put(80,20){\circle*{5}}
\put(90,20){\circle*{5}}
\put(100,20){\circle{5}}
\put(70,30){\circle{5}}
\put(80,30){\circle*{5}}
\put(90,30){\circle*{5}}
\put(100,30){\circle{5}}
\put(70,40){\circle{5}}
\put(80,40){\circle*{5}}
\put(90,40){\circle*{5}}
\put(100,40){\circle{5}}
%
\put(140,10){\circle{5}}
\put(150,10){\circle{5}}
\put(160,10){\circle*{5}}
\put(170,10){\circle*{5}}
\put(140,20){\circle{5}}
\put(150,20){\circle{5}}
\put(160,20){\circle*{5}}
\put(170,20){\circle*{5}}
\put(140,30){\circle{5}}
\put(150,30){\circle{5}}
\put(160,30){\circle*{5}}
\put(170,30){\circle*{5}}
\put(140,40){\circle{5}}
\put(150,40){\circle{5}}
\put(160,40){\circle*{5}}
\put(170,40){\circle*{5}}
%
\put(210,10){\circle*{5}}
\put(220,10){\circle*{5}}
\put(230,10){\circle*{5}}
\put(240,10){\circle*{5}}
\put(210,20){\circle*{5}}
\put(220,20){\circle*{5}}
\put(230,20){\circle*{5}}
\put(240,20){\circle*{5}}
\put(210,30){\circle{5}}
\put(220,30){\circle{5}}
\put(230,30){\circle{5}}
\put(240,30){\circle{5}}
\put(210,40){\circle{5}}
\put(220,40){\circle{5}}
\put(230,40){\circle{5}}
\put(240,40){\circle{5}}
%
\put(280,10){\circle{5}}
\put(290,10){\circle{5}}
\put(300,10){\circle{5}}
\put(310,10){\circle{5}}
\put(280,20){\circle*{5}}
\put(290,20){\circle*{5}}
\put(300,20){\circle*{5}}
\put(310,20){\circle*{5}}
\put(280,30){\circle*{5}}
\put(290,30){\circle*{5}}
\put(300,30){\circle*{5}}
\put(310,30){\circle*{5}}
\put(280,40){\circle{5}}
\put(290,40){\circle{5}}
\put(300,40){\circle{5}}
\put(310,40){\circle{5}}
%
\put(350,10){\circle{5}}
\put(360,10){\circle{5}}
\put(370,10){\circle{5}}
\put(380,10){\circle{5}}
\put(350,20){\circle{5}}
\put(360,20){\circle{5}}
\put(370,20){\circle{5}}
\put(380,20){\circle{5}}
\put(350,30){\circle*{5}}
\put(360,30){\circle*{5}}
\put(370,30){\circle*{5}}
\put(380,30){\circle*{5}}
\put(350,40){\circle*{5}}
\put(360,40){\circle*{5}}
\put(370,40){\circle*{5}}
\put(380,40){\circle*{5}}
\put(15,70){\makebox(0,0){1}}
\put(155,70){\makebox(0,0){2}}
\put(225,70){\makebox(0,0){3}}
\put(295,70){\makebox(0,0){4}}
\put(365,70){\makebox(0,0){5}}
\put(15,0){\makebox(0,0){6}}
\put(85,0){\makebox(0,0){7}}
\put(155,0){\makebox(0,0){8}}
\put(225,0){\makebox(0,0){9}}
\put(295,0){\makebox(0,0){10}}
\put(365,0){\makebox(0,0){11}}
\end{picture}
\caption{\small The 11 input patterns $b_i$ on the 16 external vertices, 
represented as three basic types of $4\times 4$-matrices: $\circ$ indicates field value 0, 
$\bullet$ indicates field value $I_{\rm max}$.}
\label{fig1}
\end{figure}

The graphs used in our investigations consist of a total of $N=24$ vertices with 
$N_{\rm ext}=16$ external vertices and 
$N_{\rm int}=8$ internal vertices. The maximal 
value of $u(z,t)$ is defined to be $I_{\rm max}=10$. 
We consider 11 different input patterns $b_i$ which are shown in Fig.~\ref{fig1}. 

\begin{table}[!b] 
\begin{center}
\vspace{0.8cm} 
\begin{tabular}{r|rrrrrrrrrr} 
inputs & \multicolumn{10}{c}{attractor states $a_i$}\\ 
$B_i$   & 1  & 2  & 3  & 4  & 5  & 6  & 7 & 8  & 9  & 10 \\ \hline 
1 \    & 1  & 1  & 1  & 1  & 1  & 1  & 1  & 1  & 1  & 1 \\ 
2 \    & 2  & 2  & 2  & 2  & 2  & 2  & 2  & 2  & 2  & 2 \\ 
3 \    & 3  & 3  & 3  & 3  & 3  & 3  & 3  & 3  & 3  & 3 \\ 
4 \    & 2  & 2  & 2  & 2  & 2  & 2  & 2  & 2  & 2  & 2 \\ 
5 \    & 3  & 3  & 3  & 3  & 3  & 3  & 3  & 3  & 3  & 3 \\ 
6 \    & 4  & 4  & 4  & 4  & 4  & 4  & 4  & 4  & 4  & 4 \\ 
7 \    & 4  & 4  & 4  & 4  & 4  & 4  & 4  & 4  & 4  & 4 \\ 
8 \    & 5  & 5  & 5  & 5  & 5  & 5  & 5  & 5  & 5  & 5 \\ 
9 \    & 6  & 2  & 2  & 7  & 5  & 6  & 7  & 7  & 10 & 10 \\ 
10 \    & 8  & 8  & 8  & 5  & 5  & 8  & 5  & 8  & 5  & 5 \\ 
11 \   & 4  & 9  & 9  & 4  & 4  & 9  & 4  & 9  & 9  & 9 \\ 
\end{tabular} 
\end{center} 
\caption{\small Example of a mapping diagram for 11 inputs $B_i$ and a 
system with 10 different attractors $a_i$. The entries show the number $i$ of the 
attractor state which is obtained by applying $B_i$ (plotted vertically) to $a_i$ (plotted horizontally).} 
\label{tab0} 
\end{table} 

In order to obtain a minimal set $A$ of attractor states under which the multiplication 
of the evolution operators $B_i$ is closed, we first determine the attractor state $a_1$ 
(or states $a_i$, $i=1, ..., m$ for a limit cycle of period $m$)    
corresponding to input $B_1$, starting from a random initial distribution 
of states in $F_X$. Subsequently, $B_2$ is applied to $a_1$, and so on until 
$B_{11}$ provides the final attractor state(s). 

Next we apply all evolution operations 
$B_1,\ldots,B_{11}$ to the obtained set of attractors until no new attracting states 
are generated. The resulting set $A$ can be represented in terms of a mapping diagram. 
An example for such a mapping diagram with a relatively small
number of 10 attractor states, which are all fixed-point 
attractors, is shown in Tab.~\ref{tab0}. The corresponding field
values on the eight internal vertices are listed in Tab.~\ref{tab0a}, and
the corresponding adjacency matrix of the graph is given in Tab.~\ref{tab0b}.

\begin{table}[h] 
\begin{center} 
\begin{tabular}{r|rrrrrrrrrr} 
attractor  & \multicolumn{10}{c}{\ \ field values on internal vertices}\\ 
states $a_i$  & 1  & 2  & 3  & 4  & 5  & 6  & 7 & 8  \\ \hline 
1 \    & 0  & 0   & 0  & 0  & 0  & 0  & 0  & 0   \\ 
2 \    & 0  & 10  & 0  & 0  & 0  & 10 & 0  & 0  \\ 
3 \    & 0  & 10  & 0  & 0  & 0  & 0  & 0  & 10   \\ 
4 \    & 10 & 0   & 0  & 10 & 0  & 0  & 0  & 0   \\ 
5 \    & 10 & 0   & 0  & 0  & 0  & 0  & 10 & 0   \\ 
6 \    & 5  & 5   & 0  & 0  & 0  & 5  & 5  & 0   \\ 
7 \    & 9  & 1   & 0  & 0  & 0  & 1  & 9  & 0   \\ 
8 \    & 9  & 1   & 0  & 0  & 0  & 0  & 9  & 1   \\ 
9 \    & 9  & 1   & 0  & 9  & 0  & 0  & 0  & 1   \\ 
10 \   & 8  & 2   & 0  & 0  & 0  & 2  & 8  & 0  \\ 
\end{tabular} 
\end{center} 
\caption{\small Configuration of field values on internal vertices 
for the 10 attractors $a_i$ of Tab.~\ref{tab0}.}
\vspace{0.5cm}
\label{tab0a} 
\end{table} 

\begin{table}[!h] 
\begin{center} 
\begin{tabular}{cccccccccccccccccccccccc} 
 \multicolumn{24}{c}{$Ad(x,y)=0$ for $x \le 16$}\\  
0 & 0 & 0 & 0 & 0 & 0 & 0 & 0 & 0 & 0 & 0 & 0 & 0 & 0 & 0 & 0 & 
0 & 0 & 0 & 1 & 0 & 0 & 1 & 0 \\
0 & 0 & 0 & 0 & 0 & 0 & 0 & 0 & 0 & 0 & 0 & 0 & 0 & 0 & 0 & 0 & 
0 & 0 & 0 & 0 & 1 & 1 & 0 & 1 \\
1 & 0 & 0 & 1 & 1 & 0 & 1 & 1 & 1 & 0 & 1 & 0 & 0 & 1 & 1 & 0 & 
0 & 0 & 0 & 0 & 1 & 1 & 0 & 0 \\
0 & 1 & 0 & 0 & 0 & 0 & 0 & 0 & 0 & 1 & 0 & 0 & 1 & 1 & 0 & 1 & 
0 & 1 & 0 & 0 & 1 & 0 & 1 & 1 \\
0 & 1 & 1 & 0 & 0 & 1 & 0 & 1 & 0 & 1 & 1 & 0 & 0 & 1 & 1 & 0 & 
0 & 1 & 1 & 1 & 0 & 0 & 0 & 1 \\
0 & 0 & 0 & 1 & 0 & 0 & 0 & 1 & 0 & 1 & 0 & 1 & 0 & 0 & 1 & 0 & 
1 & 0 & 1 & 0 & 0 & 0 & 1 & 0 \\
0 & 0 & 0 & 1 & 1 & 0 & 1 & 1 & 0 & 0 & 0 & 1 & 0 & 0 & 0 & 0 & 
0 & 1 & 0 & 0 & 0 & 1 & 0 & 1 \\
0 & 0 & 0 & 0 & 1 & 0 & 1 & 0 & 1 & 0 & 1 & 0 & 1 & 0 & 0 & 0 & 
0 & 0 & 0 & 1 & 1 & 1 & 1 & 0 
\end{tabular} 
\end{center} 
\caption{\small The adjacency matrix $Ad$ for the mapping diagram in Tab.~\ref{tab0},
with $17 \le x \le 24$ plotted vertically and $1 \le y \le 24$ plotted horizontally.
Since there are no directed lines from internal vertices to external
vertices and no lines between external vertices,  $Ad(x,y)=0$ for $x\leq16$, 
only rows $x>16$ are shown. As explained in Sec.\ \ref{sec:learn} there are 
no direct connections from the 16 external vertices to the first two internal 
vertices serving as outputs.} 
\label{tab0b} 
\end{table} 

From the mapping diagram one can deduce the multiplicative structure of 
the operations $B_i$. A simple indicator for the complexity of this 
structure is the minimal number of attractor states neccessary for the 
structure to close. For the structure corresponding to mapping 
diagram in Tab.~\ref{tab0} one can see that the first eight inputs give rise to 
very simple relations: 
\[   B_i B_j = B_i \hspace{0.5cm} 
  {\rm for}~ i\leq 8 ~{\rm and}~ j~ 
  {\rm arbitrary}\, , \] 
representing projection operators. 
Furthermore, some of these elements are identical: 
\[  B_4=B_2 ~~,~~ B_5=B_3 ~~,~~B_6=B_7\, . \] 
The remaining three elements generate new elements of the multiplicative 
structure. Simple products of these three elements yield four relations, 
\[  B_9^2=B_9 ~~,~~ B_{10}^2 = B_{10} ~~,~~ 
  B_{11}^2=B_{11} ~~,~~  B_{10} B_{11} = B_8 \, , \] 
leaving us with five new elements. The total multiplicative structure 
contains more than 20 elements.   

\subsection{Learning on Graphs} 
\label{sec:learn}

In a very elementary way, the described graphs can be 
used to simulate simple supervised learning processes. This can be 
achieved by considering the inputs 
as stimuli to which the rest of the graph reacts in order to produce an 
optimal output. In order to define such an optimal 
output, two of the internal vertices (vertex 1 and 2 in
Tab.~\ref{tab0a}) are defined 
as output vertices, on which particular field values 
as given in Tab.~\ref{tab1} are defined as optimal. 
As we want to investigate how input from the 16 external vertices 
is processed onto the two output vertices by
the remaining six internal vertices, direct connections from 
external vertices to output vertices are excluded.

\begin{table} 
\begin{center} 
\begin{tabular}{c|cc} 
inputs & \multicolumn{2}{c}{output states} \\ 
$B_i$ & 1 & 2 \\ \hline 
1 & 0 & 0 \\ 
2 & 0 & $I_{\max}$ \\   
3 & 0 & $I_{\max}$ \\   
4 & 0 & $I_{\max}$ \\   
5 & 0 & $I_{\max}$ \\   
6 & $I_{\max}$ & 0 \\   
7 & $I_{\max}$ & 0 \\   
8 & $I_{\max}$ & 0 \\   
9 & $I_{\max}$ & 0 \\   
10 & $I_{\max}$ & 0 \\   
11 & $I_{\max}$ & 0    
\end{tabular} 
\end{center} 
\caption{\small Optimal output states for all inputs.}   
\label{tab1} 
\end{table} 

The learning process is intended to produce field states on the  
six internal vertices which map external vertices $B_i$ onto output states 
as close as possible to those given in Table \ref{tab1}. The internal structure 
of the graph is, thus, optimized in such a way that its links (connectivities) 
and vertices (field values) finally give rise to optimal output states. 

The following measure of variance serves to quantify the distance of actual output 
states $u(z_i)$ from optimal output states $u(z_i)_{\rm opt}$ ($i=1,2$): 
\begin{equation} 
 v = \sum_{\rm ext.~cond.} ~\sum_{t=10}^{30}~ 
     \sum_{\rm output~states} 
     (u(z_i,t)-u(z_i)_{\rm opt})^2 \, . 
\end{equation} 
The sum extends over all 11 external conditions, over 20 time 
steps (beginning after the first 10 transient time steps), and over the two 
output vertices. A variance $v=0$ implies optimal learning, i.e.~an optimized 
structure of the six internal vertices has been reached. For a random 
graph, $v$ is of the order of 15--20 $\times 10^4$. (Note that the fitness of the 
graph is related to the inverse of its variance $v$.)    

In order to find an optimized graph (an ``optimal learner'') 
with respect to a predefined input-output pattern, a random graph is 
used as an initial condition and randomly selected single-link changes 
(insertion or deletion of a directed link) are offered 
successively, implying changes of the state values 
on internal and output vertices according to Eqs.~(6) and (7).
(Note that this strategy differs from optimization based on changing the 
strength of links, e.g.~by Hebb's rule.) 
The initial random graph contains 
only undirected links, and there are no connections of 
input-input, input-output, and output-output vertices.

If the variance of a graph after a link change decreases, it is 
accepted, otherwise rejected. In this way, a sequence of graphs 
is generated with improving output behavior. In many cases the 
sequence terminates with a variance much larger than 0 (between 
$10^2$ and $10^4$). In such cases the evolution of the graph 
ends in a local minimum far away from optimal behavior. 
In other cases the sequence ends with an optimal learner, 
$v=0$. 

\section{Non-Commutativity of Inputs and \\ Non-Monotonic Complexity of Learning} 

\subsection{Output Dependence on the Sequence of Inputs}

The inputs $B_i$ for the learning process 
are always presented in the same sequence from $i=1$ up to $i=11$. 
Each input is presented for 30 time steps, after which 
the next input follows. Except the random initialization 
of the fields on the internal and output vertices at the beginning 
of the learning run, there is no randomization when 
inputs are changed. In this case, the field values 
start with the attractor state of the previous input. 

It turns out that 
the graphs not only learn to provide optimal outputs for individual 
inputs, but they learn to do so for particular sequences 
of inputs. In most cases, input $i+1$ is recognized correctly 
(in the sense that the fields on the output vertices assume the 
optimal values) only if the previous 
input $i$ was recognized correctly {\it and} the 
starting configuration of the fields for input $i+1$ corresponds to the 
attractor for input $i$.    

The multiplicative structure introduced above expresses how sensibly 
the reaction of graphs to the presentation of an input depends on previous inputs.   
For ``perfect learners'', optimally recognizing 
each input {\em independently} of the previous configuration, 
the multiplicative structure of the inputs is quite trivial: for 
any initial state, each input operation $B_i$ simply projects 
the system onto its corresponding attractor. This gives rise 
to a mapping diagram as in Tab.~\ref{tab2}. 

\begin{table} 
\begin{center} 
\begin{tabular}{c|ccccccccccc} 
input & \multicolumn{11}{c}{attractor state $a_i$}\\ 
$B_i$ & 1 & 2 & 3 & 4 & 5 & 6 & 7 & 8 & 9 & 10 & 11 \\ \hline 
1      & 1 & 1 & 1 & 1 & 1 & 1 & 1 & 1 & 1 & 1 &  1 \\ 
2      & 2 & 2 & 2 & 2 & 2 & 2 & 2 & 2 & 2 & 2 &  2 \\ 
3      & 3 & 3 & 3 & 3 & 3 & 3 & 3 & 3 & 3 & 3 &  3 \\ 
4      & 4 & 4 & 4 & 4 & 4 & 4 & 4 & 4 & 4 & 4 &  4 \\ 
5      & 5 & 5 & 5 & 5 & 5 & 5 & 5 & 5 & 5 & 5 &  5 \\ 
6      & 6 & 6 & 6 & 6 & 6 & 6 & 6 & 6 & 6 & 6 &  6 \\ 
7      & 7 & 7 & 7 & 7 & 7 & 7 & 7 & 7 & 7 & 7 &  7 \\ 
8      & 8 & 8 & 8 & 8 & 8 & 8 & 8 & 8 & 8 & 8 &  8 \\ 
9      & 9 & 9 & 9 & 9 & 9 & 9 & 9 & 9 & 9 & 9 &  9 \\ 
10    & 10 & 10 & 10 & 10 & 10 & 10 & 10 & 10 & 10 & 10 & 10 \\ 
11    & 11 & 11 & 11 & 11 & 11 & 11 & 11 & 11 & 11 & 11 & 11 \\ 
\end{tabular} 
\end{center} 
\caption{\small The mapping diagram for a perfect learner with 
11 inputs $B_i$ and 11 attractors $a_i$. The entries 
show the number $i$ of the attractor state which is obtained by 
applying $B_i$ (plotted vertically) to $a_i$ (plotted horizontally).} 
\label{tab2} 
\end{table} 

The multiplicative structure associated with Tab.~\ref{tab2} consists 
of the 11 elements $B_i$ which are idempotent, 
\begin{equation} 
\label{eqidemp} 
   B_i^2 = B_i ~~\mbox{for all } i \, , 
\end{equation} 
and satisfy the relation 
\begin{equation} 
\label{eqproj} 
   B_i B_j = B_i ~~\mbox{for all } i,j \, , 
\end{equation} 
hence they are non-commutative, though associative: 
\begin{equation} 
\label{eqass} 
   B_i ( B_j B_k ) = ( B_i B_j ) B_k = B_i ~~\mbox{for all } i,j,k \, . 
\end{equation} 

Since the optimal reaction of a graph to an input is not uniquely 
related to that input, the attractor providing an optimal output can 
be identical for different inputs. Therefore, the multiplicative structure
of input operations can be even simpler in the sense that some of the 
attractors are identical. Table~\ref{tab0} shows a corresponding example with  
less than 11 attractors. 

Deviations from Eq.~(\ref{eqproj}) indicate 
a more complicated structure of learning operations. If the elements in the 
same row (i.e.~for the same input) of the 
mapping diagram differ from each other, the reaction of the graph with 
respect to an input depends on the previous input. 
This means that the result of a learning process depends on the 
sequence in which successive learning steps are carried out. 
This implies that the multiplicative structure of input 
operations deviates from Eq.~(\ref{eqproj}). Since
the $B_i$ are mappings, associativity is valid trivially. 
However, the structure will generally be non-commutative, 
\begin{equation} 
   B_i B_j \neq B_j B_i \, , 
\end{equation} 
although it may happen that particular inputs commute, for instance when
they project onto the same attractor, such as $B_2$ and $B_4$, or
$B_3$ and $B_5$, or $B_6$ and $B_7$ in Tab.~\ref{tab0}.

We can now understand how an optimal learner differs from
a perfect learner, which recognizes inputs independently of the sequence of
their presentation. Comparing Tabs.~\ref{tab0a} and \ref{tab1} shows that
attractor $a_1$ leads to the optimal output (field values on the first 
two vertices) for input $B_1$, attractors $a_2$ and $a_3$ yield 
the optimal output for inputs $B_2-B_{5}$, and attractors $a_4$ and $a_5$
yield the optimal output for inputs $B_6-B_{11}$. In these cases, optimal
learning coincides with perfect learning. 

From Tab.~\ref{tab0} we see that inputs $B_1-B_8$ are recognized
independently of previous inputs. By contrast, inputs $B_9$, $B_{10}$ and $B_{11}$ are
recognized correctly only if the previous input is $B_8$, $B_9$ and $B_{10}$, respectively.
Table~\ref{tab0a} shows that attractors $a_7-a_{10}$ lead to an ``almost'' correct 
output for inputs $B_9-B_{11}$, and the output of $a_6$ differs considerably 
from any optimal output. Although these situations represent optimal learning,
they are different or even far from perfect learning.

If the attractor for a particular input does not consist of one 
single state (fixed point), but of a perodic sequence of states 
(limit cycle), idempotency \ref{eqidemp} does no longer hold. 
(Strictly speaking, this is only correct if the 
number of time steps $t$ in the mapping $B_i=\Phi_{t,b_i}$ 
and the length of the cycle have no common denominator. 
Otherwise, the attractor may 
consist of more states than can be detected by 
the mapping diagram or the set of inputs $B_i$.) 

Note that the structure of learning operations derived here is more general than 
an algebra (as conjectured by Gernert 2000). There is no 
identity element, there is no neutral element, and no addition 
of the elements $B_i$ is defined. 

\subsection{Number of Attractors Versus Variance} 

In order to investigate the evolution of the set of attractors 
during the learning process, we focus on the number $N$ of attractor states 
as a function of learning steps for the entire sequence of graphs starting from a 
random graph until a graph with optimal learning is reached.
Since a large number of attractors intuitively relates to quite complex structures
of the graph during the learning process, we propose to refer to the size of the set of 
attractors as a possible measure for the {\it complexity of learning}. However, it should be 
emphasized that a rigorous definition of complexity (cf.~Wackerbauer et al.~1994) is not yet 
associated with this notion. 
 
\begin{figure}[htb] 
\begin{center}
\epsfig{figure=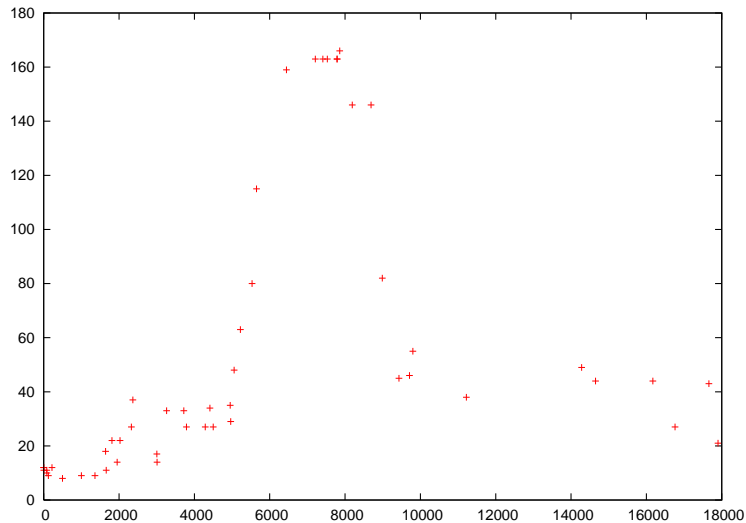,scale=0.8} 
\caption{\small Number $N$ of attractor states (vertical axis) as a function of variance 
$v$ (horizontal axis), starting from a random graph with $v\approx 18000$ 
and terminating at a graph with optimal response and $v=0$. 
The points refer to those graphs which were accepted during the 
learning process, so that decreasing variance indicates progressive
learning. The non-monotonic behavior of the complexity of learning
is clearly visible.} 
\label{fig3} 
\end{center}
\end{figure}

Initially, the graphs are (almost) random and exhibit large 
variances of the order of $2\times 10^4$. For these graphs the 
number of attractor states with respect to the inputs 
varies over a large range; typical are numbers between 
30 and 50. As learning begins, the variance decreases, but 
the number of attractor states increases, sometimes up to a few 
hundred. A further decrease in variance, below 
a value of 6000, causes the number of attractor states to decrease 
again. For optimal learners (graphs with vanishing 
variance) the number of attractor states terminates at around $N=10$.   
A typical example is shown in Fig.~\ref{fig3}. 

\begin{figure}[!htb] 
\begin{center}
\epsfig{figure=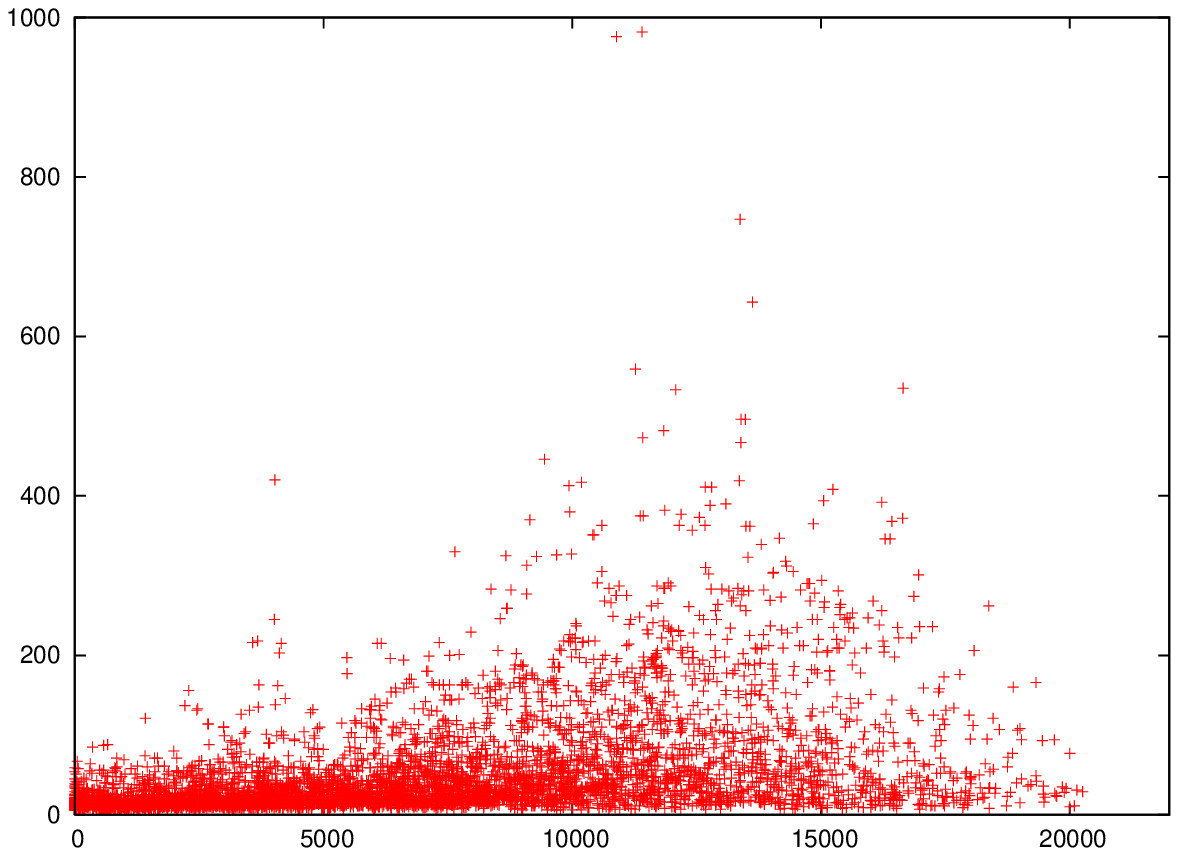,scale=0.8} 
\caption{\small Number $N$ of attractor states (vertical axis) as a function of variance $v$ 
(horizontal axis) for 116 learning sequences starting from random
graphs and terminating as (almost) optimal learners with a variance of below
10. The plot shows only those graphs which were accepted during learning.  
The non-monotonic behavior of the complexity of learning for optimal learners is clearly visible.} 
\label{fig4} 
\end{center}
\begin{center}
\epsfig{figure=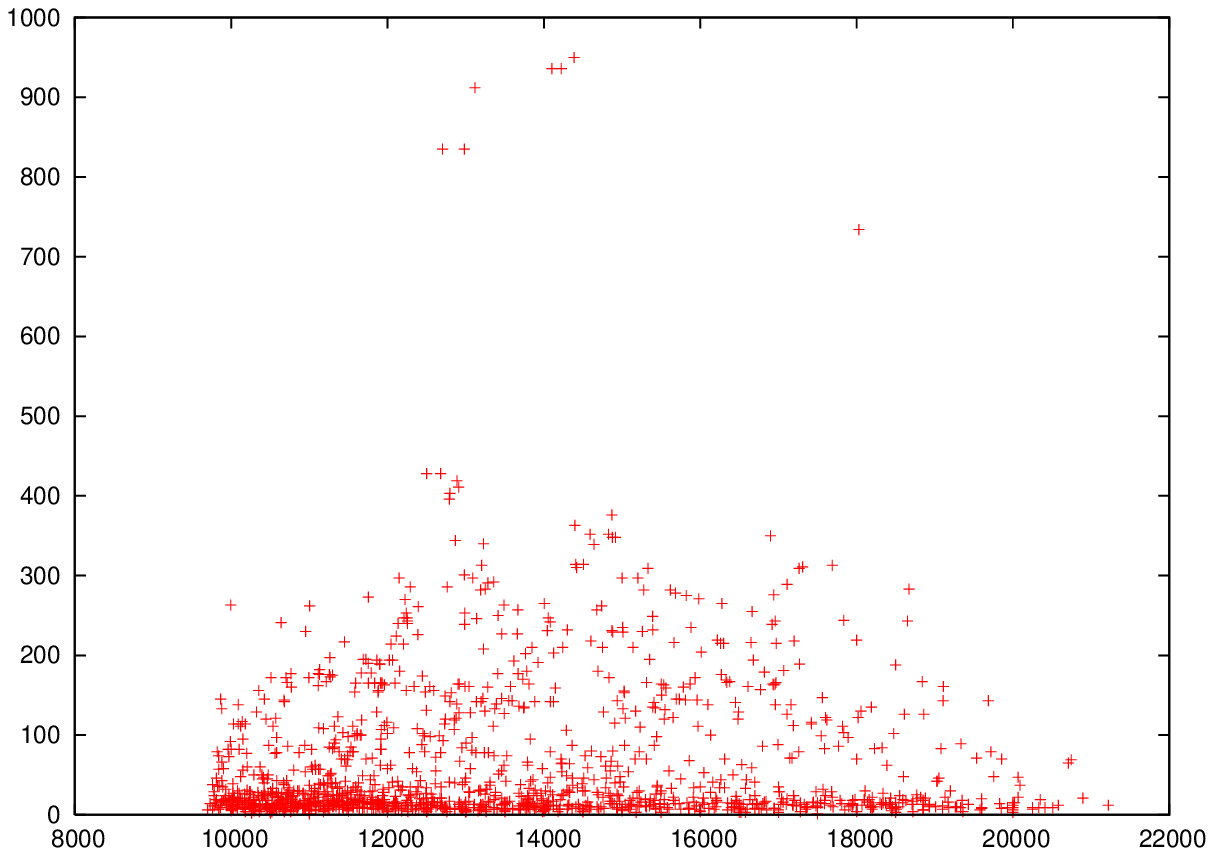,scale=0.8} 
\caption{\small Number $N$ of attractor states (vertical axis) as a function of variance $v$ 
(horizontal axis) for 98 learning sequences starting from random
graphs and terminating as non-optimal learners with a variance of above 9700.
The plot shows only those graphs which were accepted during learning.  
The non-monotonic behavior of the complexity of learning is visible for non-optimal learners 
as well.}   
\label{fig5} 
\end{center}
\end{figure}

We now select a sample of 116 learning sequences 
starting from random graphs and terminating as (almost) optimal learners. 
For this sample we count the number of attractor 
states, i.e.~the complexity of learning, for those graphs which were accepted 
during the process, i.e., for which the variance was always 
smaller than for any previous graph in the sequence.  
Their behavior can be seen in Fig.~\ref{fig4}, where $N$ is plotted 
as a function of $v$. It confirms the impression from Fig.~\ref{fig3}
that, as learning proceeds, its complexity evolves non-monotonically.

In about 50\% of the cases the sequence started with 
less than $N=50$ attractor states. The final $N$ was much smaller, 
and for intermediate stages of learning $N$ reached a maximum 
during the learning process.   
In about 85\% of all cases the final number of 
attractor configurations was smaller than 20. The 
largest final number of attractor states for an 
optimal learner was 56. 

Exceptions from this behavior occur
if the initial (random) graph has a number of attractor states 
that is extremely large, exceeding any other number of attractor 
states in the sequence. For this case we find a total number of 
15 sequences. In 12 of these sequences the initial 
number of attractors is larger than 100 (with a maximum of 747). 

Figure~\ref{fig5} shows a plot of number of attractors as a function of 
variance for 98 non-optimal learners whose final variance is 
$v > 9700$. Keeping in mind that decreasing variance corresponds to 
progressive learning, the general trend of Figs.~\ref{fig3} and \ref{fig4} reappears: 
the size of the set of attractors, i.e.~the complexity of learning, 
evolves non-monotonically as learning proceeds. 

As the main observation of the present subsection, we can state 
that the number $N$ of attractors required to optimally map a given input
onto a predetermined output evolves non-monotonically during the
process of learning. While $N$ increases during the initial phase of 
learning, it decreases again until the learning process is terminated. We interpret 
this behavior as a non-monotonic complexity of the learning process.
    
\section{Is the Complexity of Learning \\ Related to Meaning?} 

Non-monotonic as opposed to monotonic measures of complexity have 
been developed and investigated for about two decades; for a comparative overview
see Wackerbauer et al.~(1994). The property of monotonicity is usually
understood as a function of (some measure of) randomness of the pattern
or process considered. Monotonic complexity essentially increases 
as randomness increases: most random features are also most complex.
Non-monotonic complexity shows convex behavior as a 
function of increasing randomness: highest complexity is assigned to
features with a mixture of random and non-random elements, while both very 
low and very high randomness yield minimal complexity.    

\begin{figure}[!htb]
\begin{center}
\epsfig{figure=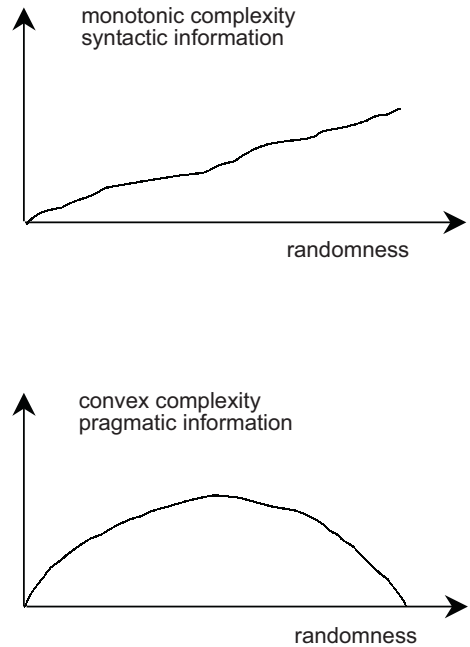,scale=0.7}
\vspace{-8cm}
\caption{\small Schematic illustration of two different classes of complexity
measures, corresponding to different information measures and distinguished by their
 functional dependence on randomness.}
\label{fig6} 
\end{center}
\end{figure}

There is an interesting relationship between the two classes of complexity
measures and measures of information; for more details see Atmanspacher (1994)
or Atmanspacher (2005). It turns out that monotonic complexity usually corresponds
to syntactic information, whereas non-monotonic (convex) complexity corresponds to
semantic information or other measures of meaning (see Fig.~\ref{fig6}).

As a particularly interesting approach, pragmatic information has been proposed
(Weizs\"acker 1972)
as an operationalized measure of meaning. Its essence is that purely random 
messages keep providing complete novelty (or primordiality) 
as they are delivered, while purely non-random messages keep
providing complete confirmation (after initial transients). 
Pragmatic information refers to meaning in terms of a mixture of 
confirmation and novelty. Extracting meaning from a message depends
on the capability to transform novel elements into knowledge using confirming elements.

It has been speculated (Atmanspacher 1994) that systems having this capacity
are able to reorganize themselves in order to flexibly modify their complexity relative to 
the task that they are supposed to solve. A learning process, in which 
insight is gained and meaning is understood, may start at low complexity (high randomness, 
much novelty) and terminate at low complexity (high regularity, much confirmation),
but it passes through an intermediate stage of maximal complexity.

The notion of pragmatic information was earlier utilized in this sense for 
non-equilibrium phase transitions in multimode lasers (Atmanspacher and Scheingraber 1990).
It could be shown that a particular well-defined type of pragmatic information, adapted
to that case, behaves precisely as indicated above. Pragmatic information is
maximal at the unstable stage of the phase transition, and it is low in the preceding
and successive stages. However, 
lasers are physical systems, and it is problematic to ascribe something like
an ``understanding of meaning'' to their behavior. 

Biological networks such as studied in this 
paper are more realistic systems for a concrete demonstration of the basic idea.  
The non-monotonic complexity of learning processes as indicated in Sec.~3.2
starts with random graphs and ends with graphs of minimized variance (maximized fitness), 
which are as non-random as possible under the given conditions. In this sense, a scenario
has been established
in which the complexity of learning on graphs qualitatively satisfies the conditions required 
for relating it to a measure of pragmatic information. Within this scenario, our approach 
suggests that the actual ``release of meaning'' during learning 
does not occur when the output is optimized but rather
when the complexity is maximized.

It is a long-standing desideratum to identify meaning-related physiological features in the brain 
(Freeman 2003). Since learning is a key paradigm in which the emergence of meaning can be studied, 
we hope that our approach may offer a useful perspective for progress concerning
this problem.          

\section{Summary} 

In this contribution an example of supervised learning in recurrent networks
of small size implemented on graphs is studied numerically. The elements of the 
network are treated as vertices of graphs and the connections among the elements are 
treated as links of graphs. Eleven inputs and two outputs are predefined, and the learning
process within the remaining six internal vertices is carried out such as to minimize
the difference between the actual output and the predetermined output. Optimization
of outputs is achieved by stable configurations at the internal vertices that can be 
characterized as attractors.  

Two particular features of the learning behavior of the network are investigated
in detail. First, it is shown that, in general, the mapping from inputs to outputs depends
on the sequence of inputs. Thus, the associative multiplicative structure of input operations 
represented by sets of attractors is, in general, non-commutative. Second, the size
of the set of attractors changes as the learning process evolves. With increasing 
optimization (fitness), the number of attractors increases up to a maximum and then 
decreases down to a usually small final set for optimal network performance.

Assuming that the size of the set of attractors indicates the complexity of learning,
its non-monotonic behavior is of special interest. Since non-monotonic measures of 
complexity can be related to pragmatic information as a measure of meaning, it is 
tempting to consider the maximum of complexity as reflecting the release of meaning in
learning processes. Further work will be necessary to substantiate this speculation.

\section*{References} 

\begin{description} 

\item Atmanspacher, H., 1994, Complexity and meaning as a bridge across the
Cartesian cut. Journal of Consciousness Studies 1, 168--181. 

\item Atmanspacher, H., 2005, A semiotic approach to complex systems, in 
Aspects of Automatic Text Analysis, ed. by A. Mehler and R. K\"ohler, 
Springer, Berlin, pp. 67--79. 

\item Atmanspacher, H., Filk, T., and Scheingraber, H., 2005, Stability analysis 
of coupled map lattices at locally unstable fixed points, European Physical 
Journal B 44, 229--239.   

\item Atmanspacher, H., and Scheingraber, H., 1990, Pragmatic information and dynamical
instabilities in a multimode continuous-wave dye laser, Canadian Journal of Physics 
68, 728--737.   

\item Bornholdt, S., and Schuster, H.G., eds., 2003,  Handbook of Graphs and Networks, 
Wiley-VCH, Weinheim. 

\item Dayan, P., and Abbott, L.F., 2001, Theoretical Neuroscience: Computational 
and Mathematical Modeling of Neural Systems, MIT Press, Cambridge MA.   

\item Duda, R.O., Hart, P.E., and Stork, D.G., 2000, Pattern Classification,
Wiley, New York.

\item Freeman, W.J., 2003, A neurobiological theory of meaning in perception, part I:
Information and meaning in nonconvergent and nonlocal brain dynamics. 
International Journal of Bifurcation and Chaos 13, 2493--2511. 

\item Gernert, D., 1997, Graph grammars as an analytical tool in physics and biology. 
BioSystems 43, 179--187. 

\item Gernert, D., 2000, Towards a closed description of observation processes. 
BioSystems 54, 165--180. 

\item beim Graben, P., 2004, Incompatible implementations of physical symbol systems.
Mind and Matter 2(2), 29--51.

\item Haykin, S., 1999, Neural Networks: A Comprehensive Foundation. 
Prentice Hall, Saddle River NJ. 

\item Hertz, J.,  Krogh, A., and Palmer, R.G., 1991, Introduction to the Theory of 
Neural Computation, Addison-Wesley, Reading MA. 

\item Jordan, M.I., ed., 1998, Learning in Graphical Models, MIT Press, Cambridge MA. 

\item Jost, J., and Joy, M.P., 2002, Spectral properties and synchronization in coupled map lattices. 
Physical Review E,  65, 016201. 

\item Paton, R., 2002, Process, structure, and context in relation to integrative biology. 
BioSystems 64, 63--72. 

\item Paton, R., 2002, Diagrammatic representations for modelling biological knowledge. 
BioSystems 66, 43--53. 

\item Sejnowski, T., ed., 2001, Graphical Models: Foundations of Neural Computation, 
MIT Press, Cambridge MA. 

\item Smolensky, P., 1988, On the proper treatment of connectionism.
Behavioral and Brain Sciences 11, 1--74. 

\item Wackerbauer R., Witt, A., Atmanspacher, H., Kurths, J., and Scheingraber, H.,  1994, 
A comparative classification of complexity measures, Chaos, Solitons \& Fractals 4, 
133--173. 

\item Weizs\"acker, E.~von, 1974, Erstmaligkeit und Best{\"a}tigung als
Komponenten der pragmatischen Information, in {\it Offene
   Systeme I}, ed.~by E.~von Weizs{\"a}cker, Klett-Cotta, Stuttgart, pp.~83--113.

\item Wilson, R.J., 1985, Introduction to Graph Theory, 
Longman Scientific \& Technical, Essex. 

\end{description}

\end{document}